\documentclass{svjour3}                     
\smartqed  
\usepackage{graphicx}
 \journalname{Journal of Intelligent Manufacturing}
\begin{document}

\title{Coordinating metaheuristic agents with swarm intelligence
}



\author{Mehmet E. Aydin         
}


\institute{M. E. Aydin \at
              University of Bedfordshire, \\
              Dept. of Computer Science and Technologies, \\
              Luton, UK\\
              \email{mehmet.aydin@beds.ac.uk}
}


\maketitle

\begin{abstract}
Coordination of multi agent systems remains as a problem since
there is no prominent method to completely solve this problem.
Metaheuristic agents are specific implementations of multi-agent
systems, which imposes working together to solve optimisation
problems with metaheuristic algorithms. The idea borrowed from
swarm intelligence seems working much better than those
implementations suggested before. This paper reports the
performance of swarms of simulated annealing agents collaborating
with particle swarm optimization algorithm. The proposed approach
is implemented for multidimensional knapsack problem and has
resulted much better than some other works published before.

\keywords{ metaheuristic agents, \and swarm intelligence, \and
particle swarm optimization, \and simulated annealing}

\end{abstract}

\section{Introduction}
\label{intro} Metaheuristic agents are collaborating agents to
solve large scale optimisation problems in the manner of multi
agent systems in which metaheuristic algorithms are adopted by the
agents as the problem solvers.  They are multi-agent systems
identified to describe teams of search agents to operate for
optimisation. This type of multi-agent systems is specific to
implementations of metaheuristics to solve large scale
optimisation problems (Aydin 2007). Coordination of multi agent systems remains as a problem since
there is no prominent method to completely solve this problem. The-state-of-the-art of coordinating multi agents via machine learning has been extensively discussed in Panait and Luke (2005) while Vazquez-Salcada et al (2005) and Kolp et al (2006) bring forward organizational and architectural issues of multi-agent systems. Since metaheuristic agents are more specific and heavily loaded in duty, their coordination is more than those are used in modelling social problems.
The coordination problem with metaheuristic agents constitutes of the eminent problem with
metaheuristics, which is that there is no guarantee provided to
find optimum solutions within a reasonable time with any
metaheuristic algorithm. Instead, they usually provide with local
optimum, which may not be satisfactory sometimes. One way to
overcome this problem is to diversify the search conducted with
the heuristics.
On the other hand, distributed problem solving is mainly expected to
bring more simplicity and reduction in computational time and
complexity, which leads to more diversity, and more reasonable solutions.
A well studied multi agent system can tackle multiple
regions of the search space simultaneously. Multiple independent
runs of the algorithms, which offer distributing the systems over
the particular metaheuristic agents, have capabilities to carry
out concurrent search within search spaces.

In this paper, the coordination problem of multi-agent systems has
been tackled once again, but, with swarm intelligence algorithms
this time. It is observed as expected that swarm intelligence
algorithms help for better interactions and information/experience
exchange. We illustrated the idea in coordinating simulated
annealing agents with particle swarm optimisation algorithms
implemented to solve multidimensional knapsack problem. Although
there are various hybrid implementations of particle swarm
optimisation and simulated annealing to solve combinatorial
problems (Chan et al 2006;Dong and Qui 2006;Wang et al 2007), we have not come across with implementation of particle
swarm optimisation algorithms to coordinate any metaheuristic agent such as simulated annealing
agents neither any distributed versions of such hybrid algorithms. In addition,
multidimensional knapsack problem has not been tackled with such
hybrid algorithm either.

Previously, a couple of multi agent coordination approaches
applied to metaheuristic agent teams to examine their performance
in coordinating them (Aydin 2007; Hammami and Ghediera 2005).
Obviously, each one provides with different benefits in tackling
search and problem solving. However, swarm intelligence has not
been considered for this coordination problems, whereas the notion
of swarm intelligence is to substantiate artificial societies
inspiring of the natural life. That is that the individuals form
up a swarm are to be considered as particular agents. In contrary,
the individuals remain as ordinary solutions not agents enabled
with various artificial skills. In this paper, we try to prove the
concept of coordinating agents with swarm intelligence algorithms.

Multidimensional knapsack problem is one of the most tackled
combinatorial optimisation problems due to its flexibility in
convertibility into the real world problems. The problem briefly
is to maximise the total weighted p index subject to the
constraints where x is a binary variable and r is a matrix of
coefficients that is imposed to limit the capacities and b is the
vector of upper limits.
\begin{equation}
\label{obj}
Maximise \sum_{j=1}^n{p_jx_j}
\end{equation}
Subject to:
\begin{equation}
\label{cons_1}
\sum_{j=1}^n{r_{ij}x_j}\leq{b_i}             {i=(1,...,m)}
\end{equation}
\begin{equation}
\label{cons_2}
x_j\in{[0,1]}            {j=(1,...,n)}
\end{equation}
Equation (1) is the objective function which measures the overall
capacity of the knapsacks used while Equation (2) and (3) provide
the hard constraints where (2) declares the upper limit of each
knapsack and (3) makes sure that the decision variable, x, can
only take binary integer values. The knapsack problem has been inspired by many application areas such as networking problems, supply chain modeling problems etc. Wilbaut et al (2008) introduce a survey on the variety of knapsack problems and the ways to solve them.

The rest of the paper is organised as follows. The second section
is to briefly introduce the notions of metaheuristic agents and
swarm intelligence with short presentation of considered
metaheuristics within the study; they are particle swarm
optimisation (PSO), bee colony optimisation (BCO), and simulated
annealing (SA) algorithms. The third section is to describe how to
deliver the coordination of a swarm of simulated annealing agents
using BCO and PSO. The experimental results are provided in
section four following by the conclusions in section five.

\section{Metaheuristic Agents and Swarm Intelligence}
\label{sec:1} The concept of metaheuristic agents is identified to
describe multi agent systems equipped with metaheuristics to
tackle hard optimisation problems. The idea of multi agency is to
build up intelligent autonomous entities whose form up teams and
solve problems in harmony. The agents equipped with metaheuristics
aim to solve hard and large-scale problems with their own
intelligent search skills. Since standalone heuristic search
usually face with local minima, ideas such as memetic algorithms,
hybrid algorithms etc. have received intensive attention to
overcome such shortcomings.  On the other hand, the idea of multi
agency eases building collaboration among various methods and
approaches in a form of collaborating independent computational
entities (Panait and Luke 2005;Vazquez-Salcada et al 2005;Kolp et al 2006).

Metaheuristic applications have been implemented as mostly
standalone systems in an ordinary sense and examined under the
circumstances of their own standalone systems. Few multi agent
implementations in which metaheuristics have been exploited are
examined in the literature. Various implementations of
metaheuristic agents have been overviewed with respect to
topologies and achievements in Aydin (2007) and Hammami and
Ghediera (2005).

Swarm intelligence is referred to artificial intelligence (AI)
systems where an intelligent behaviour can emerge as the outcome
of the self-organisation of a collection of simple agents,
organisms or individuals. Simple organisms that live in colonies;
such as ants, bees, bird flocks etc. have long fascinated many
people for their collective intelligence that is manifested in
many of the things that they do. A population of simple units can
interact with each other as well as their environment without
using any set of instruction(s) to proceed, and compose a swarm
intelligence system.

The swarm intelligence approaches are to reveal the collective
behaviour of social insects in performing specific duties; it is
about modelling the behaviour of those social insects and use
these models as a basis upon which varieties of artificial entities can be
developed. In such a way, the problems can be solved by models
that exploit the problem solving capabilities of social insects. The
motivation is to model the simple behaviours of individuals and
the local interactions with the environment and neighbouring
individuals, in order to obtain more complex behaviours that can
be used to solve complex problems, mostly optimisation problems
(Colorno et al 1994; Kennedy and Eberhart 1995; Tasgetiren et al
2007).

\subsection{Bee colonies}\label{sec:1.1} Bee colonies are rather recently
developed sort of swarm intelligence algorithms, which are
inspired of the social behaviour of bee colonies. This family of
algorithms has been successfully used for various applications
such as modelling oh communication networks (Farooq 2008), manufacturing cell formation (Pham et al 2007), training artificial neural networks (Pham et al 2006). There is a rather
common opinion on that bee colony algorithms are more successful
in continuous problems than combinatorial problems.  The main idea
behind a simple bee colony optimisation algorithm is to follow the
most successful member of the colony in conducting the search. The
scenario followed is that once a bee found a fruitful region, then
it performs the waggle dance to communicate to the rest of the
colony. Once any member of the colony realises that there is a
waggle dance performance by a peer fellow, then it moves to that
member's neighbourhood to collect more food. Inspiring of this
natural process, bee colony optimisation algorithms are
implemented for efficient search methodologies borrowing this idea
to direct the search to a more fruitful region of the search
space. That would result a quicker search for an appropriate
solution to be considered as a neat near-optimum.  For further
information Pham et al (2006), (2007) and Farooq (2008) can be seen.

\subsection{Particle swarm optimisation (PSO)} \label{sec:1.2}PSO is a
population-based optimization technique inspired of social
behaviour of bird flocking and fish schooling. PSO inventors were
implementing such scenarios based on natural processes explained
below to solve the optimization problems. Suppose the following
scenario: a group of birds are randomly searching for food in an
area, where there is only one piece of food available and none of
them knows where it is, but they can estimate how far it would be.
The problem here is "what is the best way to find and get
that food". Obviously, the simplest strategy is to follow the bird
known as the nearest one to the food. In PSO, each single
solution, called a particle, is considered as a bird, the group
becomes a swarm (population) and the search space is the area to
explore. Each particle has a fitness value calculated by a fitness
function, and a velocity of flying towards the optimum, food. All
particles search across the problem space following the particle
nearest to the optimum. PSO starts with initial population of
solutions, which is updated iteration-by-iteration.

The pure PSO algorithm builds each particle based on, mainly, two
key vectors; position $\mathbf{x_i}$, and velocity $\mathbf{v_i}$. Here,
$\mathbf{x_i} = \{x_{i1},...,x_{in}\}$, denotes the $i^{th}$
position vector in the swarm, where $x_{ik}$, is the position
value of the $i^{th}$ particle with respect to the $k^{th}$
dimension $(k=1,2,3,…,n)$, while $\mathbf{v_i} =
\{v_{i,1},...,v_{i,n}\}$ denotes the $i^{th}$ velocity vector in
the swarm, where $v_{ik}$ is the velocity value of the $i^{th}$
particle with respect to the $k^{th}$ dimension. Initially, the
position and velocity vectors are generated as continuous sets of
values randomly uniformly. Personal best and global best of the
swarm are determined at each iteration following by updating the
velocity and position vectors using :
\begin{equation}
v_{ik}(t+1) = \delta (w_t v_{ik}(t)+c_1
r_1(y_{ik}(t)-x_{ik}(t))+c_2 r_2(g_k(t)-x_{ik}(t))) \label{update}
\end{equation}
where $w$  is the inertia weight used to control the impact of the
previous velocities on the current one, which is decremented by
$\beta$, decrement factor, via $w_{t+1}=w_{t}\times \beta $,
$\delta$ is constriction factor which keeps the effects of the
randomized weight within the certain range. In addition, $r_1$ and
$r_2$ are random numbers in [0,1] and $c_1$ and $c_2$ are the
learning factors, which are also called social and cognitive
parameters. The next step is to update the positions in the
following way.
\begin{equation}
x_{ik}(t+1)  =  x_{ik}(t)+v_{ik}(t). \label{final_update}
\end{equation}

After getting position values updated for all particles, the
corresponding solutions with their fitness values are calculated
so as to start a new iteration if the predetermined stopping
criterion is not satisfied. For further information, Kennedy and
Eberhart (1995) and  Tasgetiren et al (2007) can be seen.

PSO has initially been developed for continuous problems not for
discrete ones. As MKP is a discrete problem, we use one of
discrete PSO, which is proposed by Kennedy and Eberhart (1997).
The idea is to create a binary position vector based on velocities as follows:
\begin{equation}
x_{ik}(t+1)  =  \frac{1}{e^{v_{ik}(t+1)}}.
\label{final_update_1}
\end{equation}

where equation (5) is replaced with (6) so as to produce binary
values for  position vectors.

\subsection{Simulated annealing}\label{sec:1.3} Simulated annealing (SA) is one of
the most powerful metaheuristics used in optimisation of many
combinatorial problems, which relies on a stochastic decision
making process in which a control parameter called temperature is
employed to evaluate the probability of moving within the
neighbourhood of a particular solution. The algorithm explores
across the whole search space of the problem undertaken throughout
a simulated cooling process, which gradually cools a given initial
hot temperature to a predefined frozen level. Given a search space
$S$, and a particular state in search space, $x\in{S}$, a
neighbourhood function, $N(x)$, conducts a move from $x$, to
$\acute{x}\in{S}$, where the decision to promote the state is made
subject to the following stochastic rule:-
\begin{eqnarray}
x_{i+1}=\left\{\begin {array}{lr}
\acute{x}_i&\textrm{$\Delta x > 0$}\\
\acute{x}_i&\textrm{$e^{\frac {-\Delta x}{t_i}}\geq\rho$}\\
x_i&\textrm{otherwise}
\end{array}\right.
\end{eqnarray}

where $\Delta{x}=\acute{x_i}-x_i$, $i$ is the iteration index,
$\rho$ is the random number generated for making a stochastic
decision for the new solution and $t_i$ is the level of
temperature (at the $i^{th}$ iteration), which is controlled by a
particular cooling schedule, $f(t_i)$.  This means that, in order
to make the new solution, $\acute{x_i}$, qualified for the next
iteration, either the arithmetic difference, $\Delta{x}$, needs to
be negative or the probability determined with $e^{-\Delta{
x}/{t_i}}$ is required to be higher than the random number
generated, $\rho$, where the probability is decayed by cooling the
temperature. Every state qualified to the next iteration as the
consequence of the abovementioned stochastic rule gives away to a
perturbation in which the solution state can be refreshed and
diversified to prevent the possible local optima. A predefined
number of moves attempted in this stage are repeated per iteration
so as to stabilise cooling the temperature. Obviously, the
stochastic rule does not allow only promoting the better
solutions, but also the worse ones. However, since the probability
of promoting a worse state exponentially decays towards zero, it
is getting harder to exploit the perturbation facility in advanced
stages of this process. That is because the temperature approaches
zero as the number of iterations goes higher. More details can be
found in literature such as Kolonko (1999), Aydin and Fogarty
(2004) and Hammami and Ghediera (2005).

\section{SA agents collaborating with swarm intelligence} \label{sec:2}
As explained above, simulated annealing (SA) is one of the most commonly used metaheuristic approaches that offer a stochastic problem solving procedure. It is used for numerous and various successful applications (Kolonko 1999; Aydin and Fogarty 2004) in
combinatorial and real optimisation domains. However, it is realised that the performance of implementations significantly depend on the neighbourhood structure as well as the hardness of the problem. In order to avoid poor performance due to such reasons, SA has
been either hybridised with other peer metaheuristic algorithms such as genetic algorithm or parallelised.  The main problem remains as the diversification of the search in one way or another. In this study, agents enabled with simulated annealing algorithm are used and named as SA agents.

The original idea of swarm intelligence is to form up populations of enabled individuals for collaboratively problem solving spurposes. However, due to computational complexity and the hardship in furnishing the enabled individuals with multiple advanced functionalities, swarms are usually designed as population of individual static solutions evolved with various genetic and/or heuristic operators/algorithms. In this study, individuals forming up the swarms are agentified with various advance functionalities such as problem solving and communicating independently.The idea is cultivated as follows: a population of agents is created and developed with a search skill operating in the way of simulated annealing algorithm. Then, the population is organised to team up a swarm to solve the problems with their search functionalities and interaction abilities. Previously, SA agents have been organised in a variety of fashions such as with hill climbing algorithm or metropolis rule (Aydin and Fogarty 2004; Aydin 2007). The idea was to build a way of collaboration through system architecture, and gained a slight improvement in performance.

This study has aimed to find out a better way of organising agents in a more proactive collaboration so that the agents are to be enabled with contributing problem solving whilst coordinating. For this purposes, few algorithms have been examined; evolutionary simulated annealing, bee colony optimization and particle swarm optimization algorithms. Evolutionary simulated annealing is the one examined earlier for a similar purpose, to solve some other combinatorial problems (Aydin and Fogarty 2004; Yigit et al 2006; Kwan et al 2009) in which a population of solutions is created and then evolved with a fast-track simulated annealing operator on generation basis. It imposes that once an individual solution is operated by an SA, the resulting new solution is replaced with the old one. On the other hand, bee colony optimisation algorithm applies waggle dance principle of bee colonies in which the best found solution is given to every agent to kick-off a fresh search around the most promising neighbourhood. The resulted solutions are counted and sorted accordingly, and the best of them is chosen for the next generation. Ultimately, the third examined algorithm , which is found as the most promising method, is particle swarm optimization algorithm. It considers a swarm of SA agents interacting in the way of particle swarm optimisation algorithm operating.

Figure 1 sketches the progress of searching for optimum solution through generations reflecting how each agent plays its role and how the collaboration algorithm merges the intelligence produced by each agent. First of all, a swarm of SA agents is created, where each agent starts searching with a randomly generated problem state, $\mathbf{x_i}(0)$. Once they finish a single run, the improved solutions, $\mathbf{x'_i}(0)$, are collected into a pool and applied with a particular collaboration algorithm for exchanging information purpose. This step puts very significant impact on the speed of approximation with which the collected solutions are operated with a second algorithm to exchange information for further steps, which helps the search with diversification. There, whichever algorithm is operating will shake up and reshuffle the set of solutions, and as a result the diversifications will be re-cultivated each time. This brings an easy way of switching to different neighbourhoods within the search space. This procedure continues until a pre-defined criterion is satisfied, which is indicated in Figure 1 as the termination state of the process. The final set of results, $\mathbf{x'_i}(t)$, are merged into the final pool, and a near optimum is finally determined.

The interaction of the SA agents in this way reminds the idea of variable neighbourhood search (Hansen et al 2004; Sevkli and Aydin
2006) where a systematic switch-off between search algorithms is organised in order to diversify the solutions. In an overall point
of view, the swarm of SA agents sounds borrowing this idea to implement it in a wider context of exploration.

\begin{figure}
  \includegraphics[width=12cm]{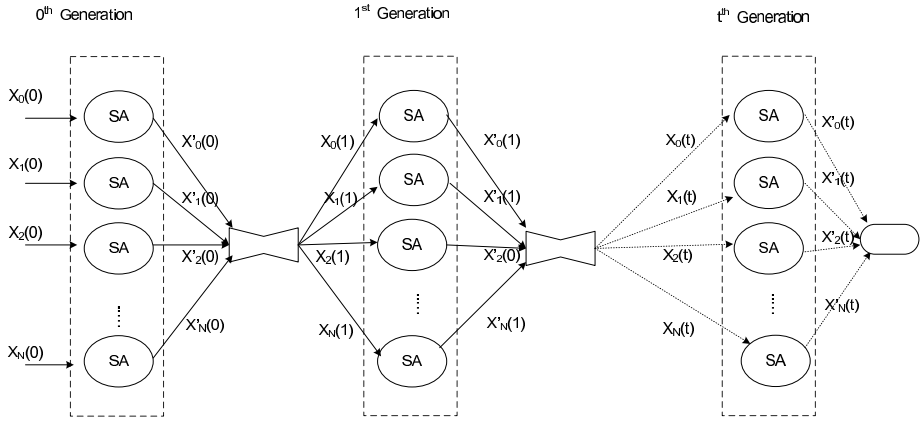}
  \caption{The progress of evolving solutions through a typical swarm of SA agents}
  \label{fig:1}
\end{figure}


The multidimensional knapsack problem is represented in a binary way to be inline with the integer programming model in which a decision variable of $\mathbf{x} = \{x_{1},...,x_{K}\}$  plays the main role in process of optimisation, where $\mathbf{x} $ is a vector of $K$ binary variables. This is also the way how to present a problem state. Here, once a corresponding amount is decided to be included in knapsack $k$, then $x_{k}$ becomes $1$ otherwise $0$. The heuristic search for optimum value is conducted via use of
neighbourhood structure of inverter function, which simply inverts the value of a randomly selected  variable at a time. The main
search is conducted by a so-called fast-track SA algorithm embedded in each agent with inverting values of up to 3 variables
at a time. A complete search operation by a SA agent is measured based a cost/fitness function, which relates each state of the
problem to a corresponding real value. 
\begin{equation}
f_i:\mathbf{x_i(t)}\longrightarrow \Re \label{fitness}
\end{equation}
where $\mathbf{x_i}$ is the $i^{th}$ vector of decision variables within the swarm, which corresponds to the $i^{th}$ SA agent.
In the case of multidimensional knapsack problem, the fitness/cost function, $f_i$, corresponds to the objective function (Equation (\ref{obj})). An agent embedded with fast-track SA explores for better state of the problem taking $\mathbf{x}_i^h = \mathbf{x}_i(t)$ and producing $\mathbf{x}_i^f = \mathbf{x'}_i(t)$ following the main procedure of SA algorithm,

\begin{equation}
\label{sa_func}
\mathbf{x}_i^f = SA_i(\mathbf{x}_i^h)
\end{equation}
where $i$ is the index for agents,
$h$ and $f$ represent "\emph{hot}" and "\emph{frozen}" keywords \footnote{"Hot" and "frozen" are two preferred keywords to express
the "initial" and "final", respectively, in order to be inline with the jargon used in simulated annealing studies.} and $SA_i(.)$ is the problem solving process of the $i^{th}$ agent. There, the improvement towards the optimum value is measured as $f_{hot} $ to $f_{frozen} $. As expected, the overall search by the whole swarm of SA agents is conducted generation-by-generation as is done in other evolutionary methodologies. Hence, implementing these multiple SA agents, there will be $N$ number of initial states of the problem considered by $N$ agents and $N$ number of improved results produced per generation. The whole swarm will include a set of fitness values representing the state of the swarm with respect to the solution quality. $\mathbf{F(t)}=\{f_0,...,f_K \}$ is the fitness vector of generation $t$ through the overall problem solving process. The swarm of SA agents will find the best of the generation, $\mathbf{x}^b(t)$, based on the fitness vector, which provides $f_{best}$. Moving to the next generation is subject to the level of satisfaction with the solution quality. If it is not sufficiently optimised, yet, the next generation will be gone through the determination of new set of hot solutions, where a coordination algorithm is needed to combine all the experiences of the agents, and let them select their new hot states. As explained before, the coordination algorithms considered in this research are evolutionary simulated annealing (ESA), bee colony optimisation (BCO) and particle swarm optimisation (PSO). ESA imposes each agent to take up $\mathbf{x}_i^f(t)$ as $\mathbf{x}_i^h(t+1)$, where $t$ is the index for generations, while BCO imposes $\mathbf{x}^b(t)$ to every agents to kick off search for next generation. PSO runs the usual interaction procedure, which explained above, to determine the new hot solutions. Therefore, a new hot solution will be produced as the result of $\mathbf{x}_i^h(t+1)$ = $\mathbf{pso}_t(\mathbf{x}_i^f, \mathbf{x}_i^{pb}, \mathbf{x}^b)$, where $\mathbf{x}_i^{pb}$ and $\mathbf{x}^b$ are personal and global best solutions. The whole procedure of coordination by PSO lasts between $\mathbf{pso}_0(.)$ and $\mathbf{pso}_T(.)$, where $T$ is the final generation through the whole process.

\section{Experimental Study}\label{sec:3}
This experimental study is not especially to solve multidimensional knapsack problem (MKP), but to test the performance of various approaches including swarm intelligence to coordinate metaheuristic agents. The abovementioned swarm intelligence model for SA agents has been examined with solving multidimensional knapsack problem, which is one of well-known NP-Hard combinatorial optimization problems. For this purpose, a swarm of SA agents, each was configured with a fast-track SA procedure, was created. Three approaches are examined for the purpose of an efficient coordination: an evolutionary simulated annealing (ESA) algorithm (Aydin and Fogarty 2004), a bee colony optimisation (BCO) algorithm (Pham et al; 2006;2007), and a binary represented PSO algorithm (Kennedy and Eberhart 1997), were implemented to work as a coordinator algorithm. The multidimensional knapsack problem was represented with a binary coding scheme.

SA procedure to be run by each agent was investigated for whether to be a 100 iteration long SA to run through 300 generations or a 200 iteration long SA to run 300 generations. The preliminary results confirmed that a 200 iteration long SA algorithm with varying
number of generations (Aydin 2008). That was inline with previous researches. In addition, the size of swarm was investigated in a range of 5 to 50. The experimentation is conducted with only two moderately hard MKP benchmarks, namely MKP6 and MKP7 collected from OR library (Beasley 1990). The results are summarised in Table~\ref{tab:1},~\ref{tab:2} and~\ref{tab:3} with the solution quality and computational time, where the solution quality is measured with relative percentage of error (RPE).
\begin{equation}
RPE = \frac{f_{opt}-f_{avrg}}{f_{opt}}
\end{equation}
where $f_{opt}$  and $f_{avrg}$ are the optimum and the average values of experimented results. The average value, $f_{avrg}$ , is
the mean calculated over 50 replications. The second performance measure is the averaged CPU time, which is the mean of the 50
replications. The performance with respect to the solution quality is primarily considered and the one with respect to CPU is
secondarily considered in case of any tight comparisons.

The implementation of the systems has been done using POP C++, which is a GRID programming language developed by Nguyan and
Kuonen (2007). It is such a unique distributed programming language that uses object distribution over the targeted
infrastructure, and arrange automatic communications among the distributed entities. This property of POP C++ eases its use in
development of multi agent systems. All experiments were conducted on GRID infrastructure in Computer Science department of Applied University of Western Switzerland in Fribourg.

\begin{table}
\caption{Experimental results of the swarm of fast-track SA agents with single inner iteration and coordinated with various
approaches} \label{tab:1}
\begin{tabular}{cccccccc}
\hline\noalign{\smallskip}
   & Swarm Size&&  ESA &&BCO &&PSO\\
   \noalign{\smallskip}\hline\noalign{\smallskip}
      &  &RPE &CPU &RPE &CPU &RPE &CPU\\
      \noalign{\smallskip}\hline\noalign{\smallskip}
    &5   &0.03495 &0.11    &0.02808 &0.73    &0.00257 &0.84\\
    &10  &0.01183 &0.43    &0.02021 &1.29    &0.00214 &1.38\\
    &15  &0.00899 &0.86    &0.01694 &1.73    &0.00170 &2.31\\
 MKP6   &20  &0.01052 &1.08    &0.01530 &2.25    &0.00203 &2.40\\
    &30  &0.00762 &1.80    &0.01344 &2.79    &0.00098 &2.56\\
    &40  &0.00768 &1.86    &0.01226 &4.34    &0.00122 &3.67\\
    &50  &0.00633 &2.56    &0.01093 &5.28    &0.00061 &4.09\\
\hline\noalign{\smallskip}
    &5   &0.03748 &0.14    &0.04077 &0.59    &0.00307 &0.78\\
    &10  &0.02170 &0.52    &0.03270 &1.18    &0.00175 &1.30\\
    &15  &0.01528 &1.01    &0.02782 &1.57    &0.00112 &1.31\\
 MKP7   &20  &0.01407 &1.17    &0.01906 &2.10    &0.00064 &1.31\\
    &30  &0.00961 &2.34    &0.01516 &2.95    &0.00014 &0.93\\
    &40  &0.00821 &2.20    &0.01736 &4.35    &0.00030 &1.21\\
    &50  &0.00865 &2.66    &0.01979 &5.38    &0.00028 &1.15\\
\hline\noalign{\smallskip}
\end{tabular}
\end{table}
Table ~\ref{tab:1} presents experimental results with the most
fast-track SA agents coordinated with all three approaches against
various swarm sizes. The SA algorithm is configured to run 200
iterations without any inner replications, which means that the
cooling schedule allows operating once per level of temperature.
All three algorithms, ESA, BCO and PSO, are separately applied to
the same swarm of SA agents under the same circumstances. The
swarm size varies between 5 and 50 agents. The multidimensional
knapsack benchmark problems tackled are MKP6 and MKP7 in all
cases. All experiments are replicated for 50 times. The worst
level of achievement with respect to quality of solution is
delivered by BCO while PSO has the best and ESA has an
intermediate level of achievement. On the other hand, the shortest
computational time achieved by ESA while the longest one is done
by BCO and PSO is in the middle. The overall gain by PSO over BCO,
which is the worst case, remain between 90-95\% and 25-33\% by ESA.
The time-wise gain is 49\% and 31\% by ESA and PSO, respectively.
The swarm-size-wise performance is a significant too. For both
benchmarks, the size of the swarm indicates a gradual increase in
performance in all cases; the solution quality index linearly
decreases.  Another most interesting fact is that the error level
indicated by PSO is nearly about 10\% of both ESA's and BCO's levels.

Table ~\ref{tab:2} presents the results of experimentations sets
which considered 5 inner iterations per SA cycle. These results
are much better ones comparing to the single inner iteration case.
All three algorithms that coordinate fast-track SA agents, with 5
inner iterations per cycle this time, and improve their performance
gradually through the growing size of the swarm. ESA hits 100\%
achievement with 30 and 40-agent swarms, while PSO hits about 99\%
in both cases. BCO remains improving in comparison with the single
inner case, but outperformed by both ESA and PSO. The overall gain
by PSO over BCO, which is the worst case remain between 65-95\%
and 84-95\% by ESA. The gain with respect to CPU times is 82\% and
39\% by ESA and PSO, respectively.

\begin{table}
\caption{Experimental results of swarm of fast-track SA agents
with 5 inner iterations and coordinated with various approaches}
\label{tab:2}       
\begin{tabular}{cccccccc}
\hline\noalign{\smallskip}
   & Swarm Size & ESA&& BCO &&PSO\\
    \noalign{\smallskip}\hline\noalign{\smallskip}
     &   &RPE &CPU &RPE &CPU &RPE &CPU\\
        \noalign{\smallskip}\hline\noalign{\smallskip}
  &  5  & 0.00069 &0.03    &0.00182 &0.64    &0.00076 &0.70\\
    &10  &0.00031 &0.34    &0.00139 &1.21    &0.00066 &1.07\\
MKP6    &15  &0.00013 &0.32    &0.00143 &1.65    &0.00068 &1.81\\
    &20  &0.00005 &0.29    &0.00100 &1.64    &0.00042 &1.33\\
    &30  &0.00000 &0.27    &0.00090 &1.91    &0.00021 &1.08\\
    &40  &0.00000 &0.20    &0.00121 &2.73    &0.00011 &1.42\\
\noalign{\smallskip}\hline\noalign{\smallskip}
    &5   &0.00031 &0.08    &0.00190 &0.56    &0.00013 &0.24\\
    &10  &0.00009 &0.30    &0.00128 &0.92    &0.00004 &0.26\\
MKP7    &15  &0.00006 &0.32    &0.00118 &1.15    &0.00009 &0.51\\
    &20  &0.00003 &0.28    &0.00120 &1.27    &0.00009 &0.65\\
    &30  &0.00000 &0.25    &0.00078 &1.37    &0.00002 &0.57\\
    &40  &0.00000 &0.28    &0.00082 &1.59    &0.00002 &0.44\\

\noalign{\smallskip}\hline
\end{tabular}
\end{table}

Table ~\ref{tab:3} shows the experimental results of more focused
SA agents, which are replicating 10 times per step of cooling
schedule. Since this way of search is more focused, the results of
both ESA and PSO hit the optimum 100\% with swarm size of 20.
Therefore, the experimentation has not proceeded further. As the
table manifests, PSO and ESA compete each other, but outperform
BCO with respect to both quality of solution and computational
time, where the gain over BCO in terms of solution quality is
82-89\% and 82-92\% by ESA and PSO, respectively. The achievement
via CPU time is 64\% and 22\% by ESA and PSO, respectively.

\begin{table}
\caption{Experimental results of ESA agents with 10 inner
iterations and coordinated with various approaches}
\label{tab:3}       
\begin{tabular}{cccccccc}
\hline\noalign{\smallskip}
&Swarm Size& &ESA&  & BCO& &PSO  \\
\noalign{\smallskip}\hline\noalign{\smallskip}
&&RPE&CPU&RPE& CPU& RPE& CPU\\
\noalign{\smallskip}\hline\noalign{\smallskip}
&5 & 0.00027&0.09&0.00086&0.44 & 0.00029& 0.48\\
MKP6&10 &  0.00002& 0.17&    0.00063 &0.66&    0.00013 &0.66\\
&15 & 0.00000 &0.14 &   0.00066 &0.80   &0.00008 &0.49\\
&20 & 0.00000 &0.14&   0.00060 &0.97  &0.00000 &0.33\\
\noalign{\smallskip}\hline\noalign{\smallskip}
&5   &0.00072 &0.16  &0.00141 &0.45  &0.00019 &0.35\\
MKP7 &10 &0.00000 &0.13   &0.00130 &0.62 & 0.00013 &0.55\\
    &15  &0.00000 &0.13   &0.00070 &0.64 & 0.00002 &0.44\\
    &20  &0.00000 &0.14   &0.00073 &0.75  & 0.00000 &0.48\\
\noalign{\smallskip}\hline
\end{tabular}
\end{table}

\begin{figure*}[!htbp]
 \includegraphics[width=8cm, angle=270]{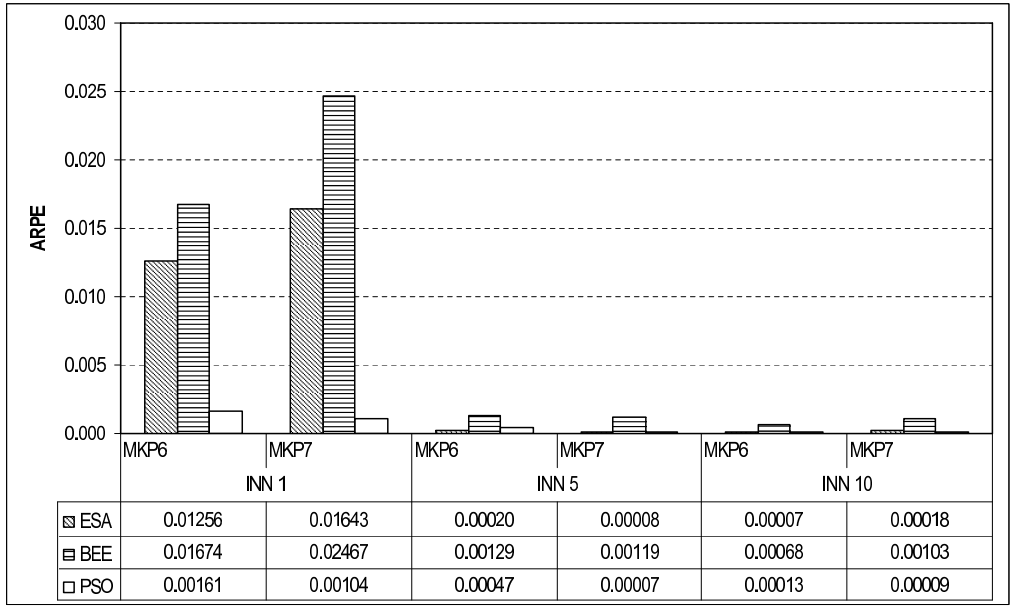}
\caption{Average performance of agent swarms in various sizes
operating with all three algorithms and all three inner-iteration
levels} \label{fig:2}
\end{figure*}

Fig.~\ref{fig:2} indicates the averaged-RPE results of each
coordinating approach per benchmark per level of inner iterations
in fast-track SA agents. The averaged results are tabulated across
horizontal axis pointing out the overall achievement of each
approach, where the benchmark problems are indicated as MPK6 and
MPK7 with each inner iteration case. INN 1, INN 5 and INN 10
indicate the inner iteration level of 1, 5 and 10. As both the
graph and the tabulated values reveal, the performance of ESA and
PSO comparable beyond the inner iterations of 5 onward. However,
their achievements remain significantly different in the case of
inner iteration 1, which is the simplest form of cooling process
in SA procedure. PSO clearly and significantly outperform both ESA
and BCO approaches, while ESA does better than BCO.  Depending on
their level of difficulty, simulated annealing algorithms are
configured with the level of inner iterations, whereas some
problems favour of higher level of inner iterations, but some do
not do at all, especially those are time sensitive such as resource
scheduling problem of radio access networks (Kwan et al 2009), where the speed of the algorithms are measured in nano-second level. Therefore, more focused and intensified search will not help solving such problems at all.

\section{Conclusions}\label{sec:4} Metaheuristic agent swarms need collaboration
in one way or another to deliver an efficient problem solving
services. In this paper, three collaboration algorithms have been
examined with respect to efficiency in solution quality. The
agents form up the swarms, which are configured as simulated
annealing agents to solve multidimensional knapsack problem.
Evolutionary simulated annealing, bee colony optimisation and
particle swarm optimisation algorithms are used for collaboration
purposes. The algorithm found best to be paired with SA agents is
PSO, which is a relatively newer swarm intelligence approach that
has good record for continuous problems, but usually needs a local
search embedded in for combinatorial problems. On the other hand
SA needs to incorporate with other search methods for
diversification. It is significantly concluded that collaborating
metaheuristic agents with swarm intelligence algorithm adds up
value into the quality of solution. This incorporation works in
the form of a variable search algorithm in an overall point of
view. It also keeps the properties of ESA (Yigit et al 2006) as it
reheats the temperature, and works with a population.

\begin{acknowledgements}
A part of this study has been carried out in  Engineering
College of Fribourg in Applied University of Western Switzerland, Fribourg,
Switzerland, while the author was visiting GRID research group
there. The author is particularly grateful to Prof Pierre Kuonen,
the head of GRID research group and Mr. Jean-Francois Roche,
senior technician of the group for their sincere and kind support
in both use of POP C++ and making use of their GRID
infrastructure. The author is also grateful to Prof. Jie Zhang
from University of Bedfordshire, Luton, UK, for his sponsorship to
the author during his visit to GRID research group.
\end{acknowledgements}

\bibliographystyle{spbasic}      


\end{document}